\documentclass[aps, prd, twocolumn, nofootinbib, floatfix, superscriptaddress]{revtex4-1}
\usepackage{subfigure}
\usepackage{graphicx}
\usepackage{dcolumn}
\usepackage{epsfig}
\usepackage{amsmath}
\usepackage{amsfonts}
\usepackage{graphicx}
\usepackage{amssymb}
\usepackage{comment}
\usepackage{amssymb,amsmath,amsfonts}
\usepackage{xfrac}
\usepackage{color}
\usepackage{xcolor}
\usepackage{tikz}

\usepackage{tcolorbox}
\usepackage{hyperref}
\hypersetup{colorlinks, citecolor=bluscuro, linkcolor=black, urlcolor=bluscuro}
\definecolor{rossos}{cmyk}{0,1,1,0.55}
\definecolor{bluscuro}{rgb}{0.15, 0.2, .85}
\definecolor{bluchiaro}{cmyk}{1,.3,0.,0.1}

\let\oldsqrt\sqrt
\def\sqrt{\mathpalette\DHLhksqrt}
\def\DHLhksqrt#1#2{%
\setbox0=\hbox{$#1\oldsqrt{#2\,}$}\dimen0=\ht0
\advance\dimen0-0.2\ht0
\setbox2=\hbox{\vrule height\ht0 depth -\dimen0}%
{\box0\lower0.4pt\box2}}

\newcommand{\sss}[1]{{\scriptscriptstyle{#1}}}

\newcommand{\uPl}{\mathrm{Pl}}

\newcommand{\usssPl}{\sss{\uPl}}

\newcommand{\Mp}{M_\usssPl}

\newcommand{\beq}{\begin{equation}}
\newcommand{\eeq}{\end{equation}}
\newcommand{\bea}{\begin{equation}\begin{aligned}}
\newcommand{\eea}{\end{aligned}\end{equation}}

\newlength{\wsingfig}
\newlength{\wdblefig}
\newlength{\wquadfig}
\newlength{\wtriplefig}
\newcommand{\Eq}[1]{Eq.~(\ref{#1})}

\newcommand{\Fig}[1]{Fig.~{\ref{#1}}}

\newcommand{\Sec}[1]{Sec.~\ref{#1}}

\newcommand{\App}[1]{Appendix~\ref{#1}}

\begin{document}

\title{Toward the primordial black hole formation threshold in a time-dependent equation-of-state background}

\author{Theodoros Papanikolaou}
\email{papaniko@noa.gr}
\affiliation{National Observatory of Athens, Lofos Nymfon, 11852 Athens, Greece}

\begin{abstract} 
\noindent
The study of the primordial black hole (PBH) gravitational collapse process requires the determination of a critical energy density perturbation threshold $\delta_\mathrm{c}$, which depends on the equation of state of the universe at the time of PBH formation.  Up to now, the majority of analytical and numerical techniques calculate $\delta_\mathrm{c}$ by assuming a constant equation-of-state (EoS) parameter $w$ at the time of PBH formation.  In this work,  after generalizing the constant $w$ prescription of ~\cite{Harada:2013epa} for the computation of $\delta_\mathrm{c}$ and commenting its limitations we give a first estimate for the PBH threshold in the case of a time-dependent $w$ background.  In particular, we apply our formalism in the case of the QCD phase transition, where the EoS parameter varies significantly with time and one expects an enhanced PBH production due to the abrupt softening of $w$. At the end, we compare our results with analytic and numerical approaches for the determination of $\delta_\mathrm{c}$ assuming a constant EoS parameter.

\end{abstract}

\maketitle

\section{Introduction}
\label{sec:intro}
PBHs, first proposed in early '70s ~\cite{1967SvA....10..602Z, Carr:1974nx,1975ApJ...201....1C}, form in the very early universe before star formation out of the gravitational collapse of very high overdensities whose energy density is higher than a critical threshold. According to recent arguments, PBHs can naturally account for a part or the totality of dark matter~\cite{Chapline:1975ojl,Clesse:2017bsw}. They can potentially explain as well the generation of large-scale structures through Poisson fluctuations~\cite{Meszaros:1975ef,Afshordi:2003zb} and seed also the supermassive black holes residing in galactic centres~\cite{1984MNRAS.206..315C, Bean:2002kx}. At the same time, PBHs are connected with numerous gravitational-wave (GW) signals like the stochastic GW background associated to black-hole merging events~\cite{Nakamura:1997sm, Ioka:1998nz, 
Eroshenko:2016hmn,Zagorac:2019ekv, Raidal:2017mfl}, like the ones recently detected by VIRGO-LIGO~\cite{LIGOScientific:2018mvr},  as well the second order GW signal induced from primordial curvature 
perturbations~\cite{Bugaev:2009zh, Saito_2009, Nakama_2015, 
Yuan:2019udt,Fumagalli:2020nvq} (for a recent 
review see \cite{Domenech:2021ztg}) or from Poisson PBH energy density 
fluctuations~\cite{Papanikolaou:2020qtd,Domenech:2020ssp,Kozaczuk:2021wcl}.  In particular, through the aforementioned GW portal PBHs can act as well as as novel probes constraining modified gravity theories~\cite{Papanikolaou:2021uhe,Papanikolaou:2022hkg}. Other hints in favor of PBHs can be found here~\cite{2018PDU....22..137C}.

In the standard PBH formation scenario,  where PBHs form out of the collapse of enhanced energy density perturbations, the PBH formation threshold $\delta_\mathrm{c}$ depends in general on the shape of the energy density perturbation profile of the collapsing overdensity region~\cite{Musco:2018rwt,Escriva:2019phb} as well as on the equation-of-state parameter at the time at which the gravitational collapse is taking place~\cite{Carr:1975qj}.  This threshold value is very important since it can affect significantly the abundance of PBHs, a quantity which is constrained by many observational probes~\cite{Carr:2020gox}. 

Historically,  the first attempt to compute the PBH formation threshold was done by B. Carr and S. Hawking between 1974 and 1975~\cite{1974MNRAS.168..399C,Carr:1975qj} where by using a Newtonian Jeans instability criterion they were led to the conclusion that $\delta_\mathrm{c}\sim w$.  Afterwards, $\delta_\mathrm{c}$ was studied  through numerical hydrodynamic simulations by  the pioneering works of Nadezhin, Novikov $\&$ Polnarev in 1978~\cite{1978SvA....22..129N}, Bicknell $\&$ Henriksen in 1979~\cite{1979ApJ...232..670B} and Novikov $\&$ Polnarev in 1980~\cite{1980SvA....24..147N} and later after a pause of 20 years by high sophisticated numerical codes by Niemeyer $\&$ Jedmazik~\cite{Niemeyer:1997mt} and Shibata $\&$ Sasaki~\cite{Shibata_1999}.

Within the last decade,  there have been a huge progress regarding the determination of the PBH formation threshold both at the analytic as well as at the numerical level.  In particular,  at the analytic level, T.Harada, C-M. Yoo $\&$ K. Kohri (HYK) in 2013~\cite{Harada:2013epa} refined the $\delta_\mathrm{c}$ value obtained by Carr in 1975 by confronting the gravitational force which pushes the fluid matter of the collapsing overdensity inwards and enhances as such  the gravitational collapse with the pressure gradient force which in general pushes the fluid outwards, thus disfavoring the collapsing process.  At the end, by comparing the time at which the pressure sound wave crosses the overdensity collapsing to a PBH with the onset time of the gravitational collapse they found that the expression for $\delta_\mathrm{c}$ in the comoving gauge reads as:
\beq\label{delta_c-HYK}
\delta_\mathrm{c}= \frac{3(1+w)}{5+3w}\sin^2\left(\frac{\pi\sqrt{w}}{1+3w}\right).
\eeq

Some years later, \cite{Escriva:2019phb,Musco:2018rwt} studied the effect of the shape of the initial energy density perturbation which collapses to a PBH on $\delta_\mathrm{c}$ by introducing the shape parameter $\alpha$ in terms of a compaction function $\mathcal{C}$ defined as
\beq\label{alpha}
\alpha \equiv -\frac{r^2_\mathrm{m}\mathcal{C}^{\prime\prime}(r_\mathrm{m})}{4\mathcal{C}(r_\mathrm{m})},
\eeq
where the compaction function $\mathcal{C}$ is defined as $\mathcal{C} \equiv 2\frac{\delta M(r,t)}{R(r,t)}$
with $\delta M(r,t)$ being the mass excess of a local overdense region and $R(r,t) = a(t)r$ being the areal radius of this region with respect to a spatially flat background metric.  The parameter $r_\mathrm{m}$ is the position where the compaction function is maximized giving in practice the characteristic scale of the collapsing overdensity while primes denote spatial derivatives\footnote{The compaction function $\mathcal{C}$ is considered here in the superhorizon regime where one can perform a gradient expansion approximation ~\cite{Musco:2018rwt} and thus is time-independent.}. This shape parameter quantifies actually the broadness or sharpness of the energy density perturbation around its peak.   In particular,  large values of $\alpha\gg 1$ correspond to a broad peak of the collapsing energy density perturbation, whereas for smaller values of $\alpha$ the energy density profile is steeper. The work of~\cite{Escriva:2019phb} was then generalized for an arbitrary EoS parameter $w$ and it was found that for $w>1/3$ one can find an analytic formula for $\delta_\mathrm{c}$ as a function of $\alpha$ and $w$.  For $w<1/3$, the determination of an analytic expression for $\delta_\mathrm{c}$ remains an open issue given that in this regime the full shape of the compaction function is necessary~\cite{Escriva:2020tak}. 

At this point, it is very important to highlight as well the immense interest raised in the recent years concerning the effect of non-linearities~\cite{Kawasaki:2019mbl,Young:2019yug,Germani:2019zez,Young:2020xmk} and non-Gaussianities~\cite{Young:2013oia,Young:2015cyn, Franciolini:2018vbk, DeLuca:2019qsy,Yoo:2019pma,Kehagias:2019eil} for the determination of the PBH formation threshold  as well as the dependence of $\delta_\mathrm	{c}$ and the PBH abundance on the details of the initial power spectrum of curvature perturbations which gave rise to PBHs~\cite{Germani:2018jgr,Yoo:2018kvb,Yoo:2020dkz,Musco:2020jjb}. In addition,  some first research works were also performed regarding the dependence of the PBH formation threshold on non sphericities~\cite{Kuhnel:2016exn,Yoo:2020lmg}, on anisotropies~\cite{Musco:2021sva} as well as on the underlying theory of gravity~\cite{Chen:2019ueb}.

All the above mentioned research works while determining $\delta_\mathrm{c}$ made the assumption that $w$ is constant in time which is a good approximation for the vast majority of the cosmic epochs~\cite{Kolb:1990vq}. However,  according to the current cosmological paradigm, the universe's EoS parameter varies with time and there are cases where someone is met with very abrupt changes in $w$ such the QCD phase transition~\cite{Philipsen:2012nu} and the (pre)reheating era~\cite{Munoz:2014eqa},  which intermediates between inflation and the Hot Big Bang (HBB) era.  In all these regimes,  a refined calculation of the PBH formation threshold is required accounting for the effect of a dynamical $w$ profile.

Furthermore, one should point out here that the associated to PBHs scalar induced stochastic gravitational wave background (SGWB) is strongly dependent on the underlying cosmological background~\cite{Domenech:2019quo,Bhattacharya:2019bvk} and can serve  as such as a probe of the thermal history of the universe~\cite{Domenech:2020kqm}. Thus,  a better understanding of the dependence of the PBH formation threshold on the EoS parameter will unavoidably entail a better understanding of the PBH formation process and the associated to it scalar induced SGWB signal potentially detected by ET ~\cite{Maggiore:2019uih},  LISA~\cite{Audley:2017drz,Caprini:2015zlo,LISACosmologyWorkingGroup:2022jok} and SKA observational probes~\cite{Janssen:2014dka}.  Interestingly,  as stated  recently in the literature, the NANOGrav signal~\cite{NANOGrav:2020bcs} can be interpreted as an induced SGWB from a close to a dust-like stage with $ -0.091<w< 0$, a result which is dependent however on the exact dependence of $\delta_\mathrm{c}$ on the dynamical profile of $w$~\cite{Domenech:2020ers}.

Therefore,  given all the above motivation regarding the effect of a dynamical $w$ profile on $\delta_\mathrm{c}$ a legitimate question to ask is how $\delta_\mathrm{c}$ is calculated in a time-dependent $w$ background? A first attempt towards this direction was performed by the pioneering work of ~\cite{Byrnes:2018clq} where $\delta_\mathrm{c}$ was computed during the QCD phase transition by making a time average of the EoS parameter between the horizon crossing time and the time of maximum expansion of the collapsing overdensity region. Then, $\delta_\mathrm{c}$ was interpolated by using tabulated data of $\delta_\mathrm{c}$ for different values of $w$ obtained from numerical simulations but under the assumption of constant $w$.  In this work, we make a first step towards a semi-analytic scheme for the computation of $\delta_\mathrm{c}$ in time-dependent $w$ backgrounds based on simple physical arguments.

Consequently,  following this introduction where we present a historic overview of the literature regarding the determination of the PBH formation threshold,  we perform in \Sec{sec:delta_c_w_t} a refined computation of $\delta_\mathrm{c}$ in a time-dependent $w$ background by generalizing the work of ~\cite{Harada:2013epa} and commenting its limitations.  Then,  in \Sec{sec:QCD}  we apply our formalism in the case of the QCD phase transition, during which $w$ varies significantly with time and PBH production is enhanced due to the softening of the EoS parameter. Finally,  \Sec{sec:conclusions} is devoted to conclusions.


\section{Mathematical formulation}\label{sec:delta_c_w_t}
In this section,  after revising the HYK prescription~\cite{Harada:2013epa} for the computation of the PBH formation threshold valid for constant $w$ we then generalize it by determining $\delta_\mathrm{c}$ accounting for the time dependence of the EoS. 

\subsection{The ``three-zone" model}
\begin{figure}[h!]
\begin{center}
\includegraphics[width=0.496\textwidth, clip=true]{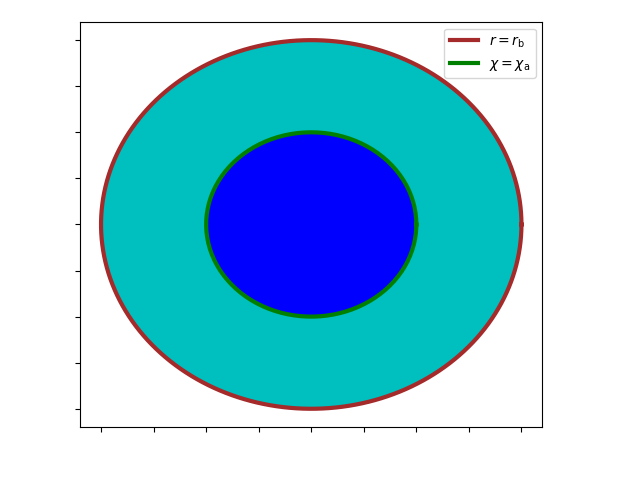}
\caption{The spherical ``three-zone" model: The overdensity region is shown in blue and it is surrounded by a spherical underdense layer displayed with cyan. The boundary between the overdensity region and the spherical underdense layer is depicted with the green circumference at $\chi=\chi_\mathrm{a}$ whereas the boundary between the underdense layer and the FLRW flat background is shown with the brown circumference at $r=r_\mathrm{b}$.}
\label{fig:three_zone_model}
\end{center}
\end{figure}

We introduce firstly the spherically symmetric ``three-zone" model where the overdense region is a homogeneous core (closed universe) surrounded by a thin underdense spherical shell which compensates the overdensity and separates the overdense region from the expanding background universe.  See \Fig{fig:three_zone_model}.

On the one hand, the background metric corresponding to a flat FLRW universe can be recast as
\beq\label{background metric}
\mathrm{d}s^2 = - \mathrm{d}t^2 + a^2_\mathrm{b}(t)\left(\mathrm{d}r^2+r^2\mathrm{d}\Omega^2\right),
\eeq
where $\mathrm{d}\Omega^2$ is the line element of a unit two-sphere and $a_\mathrm{b}(t)$ is the scale factor of the background universe. The respective Friedmann equation reads as
\beq\label{Friedmann equation - background universe}
H^2_\mathrm{b} = \left(\frac{\dot{a}_\mathrm{b}}{a_\mathrm{b}}\right)^2 = \frac{\rho_\mathrm{b}}{3\Mp^2},
\eeq
where $\rho_\mathrm{b}$ and $H_\mathrm{b}$ is the energy density and the Hubble parameter of the background.

On the other hand, the overdense region associated with a close ($K=1$) FLRW universe  is described with the following metric
\beq\label{overdense region metric}
\mathrm{d}s^2 = - \mathrm{d}t^2 + a^2(t)\left(\mathrm{d}\chi^2+\sin^2\chi \mathrm{d}\Omega^2\right)
\eeq
and the Friedmann equation reads as
\beq\label{Friedmann equation - overdense region}
H^2 = \left(\frac{\dot{a}}{a}\right)^2 = \frac{\rho}{3\Mp^2} - \frac{1}{a^2},
\eeq
where $\rho$ is the energy density of the overdense region.

The underdense spherical shell is matched to the closed FLRW universe describing the overdensity at $\chi=\chi_\mathrm{a}$ while the flat FLRW background universe is matched to the compensating underdense layer at $r=r_\mathrm{b}$. Therefore, the areal radius at the edge of the overdense region, $R_\mathrm{a}$ as well as that at the edge of the surrounding underdense spherical shell read as
\beq\label{matching conditions}
R_\mathrm{a} = a\sin\chi_\mathrm{a}, \quad R_\mathrm{b}=a_\mathrm{b}r_\mathrm{b}.
\eeq
\subsection{The PBH formation threshold in the uniform Hubble gauge}
We proceed then to the computation of $\delta_\mathrm{c}$ on the uniform Hubble gauge where the Hubble parameters of the overdensity and that of the background are the same, i.e. $H=H_\mathrm{b}$. Doing so, we define the energy density parameter $\Omega$ as 
\beq\label{Omega overdensity definition}
\Omega \equiv \frac{\rho}{3\Mp^2H^2} = 1 + \frac{1}{a^2H^2},
\eeq
where in the last equality we have used \Eq{Friedmann equation - overdense region}. Then, using the expression for the areal radius at $\chi = \chi_\mathrm{a}$, i.e.  $R_\mathrm{a} = a\sin\chi_\mathrm{a}$, as well as the definition of the horizon scale, i.e. $R_\mathrm{H}=H^{-1}$, one can find straightforwardly that 
\beq\label{Omega - sinxa relation}
(\Omega-1)\left(\frac{R_\mathrm{a}}{R_\mathrm{H}}\right)^2 = \sin^2\chi_\mathrm{a},
\eeq
an expression which relates $\Omega$ with the scale of the overdensity. In addition, one can relate $\Omega$ with the energy density contrast of the overdensity defined as
\beq\label{delta overdensity definition}
\delta \equiv \frac{\rho-\rho_\mathrm{b}}{\rho_\mathrm{b}}.
\eeq
Specifically, by solving \Eq{delta overdensity definition} for $\rho$ and substituting $\rho$ in  \Eq{Omega overdensity definition} one gets that 
\beq\label{Omega-delta relation}
\Omega = (1+\delta) \left(\frac{H_\mathrm{b}}{H}\right)^2, 
\eeq
where $\rho_\mathrm{b}$ has been expressed in terms of $H_\mathrm{b}$ through \Eq{Friedmann equation - background universe}.
Then, one can extract the energy density perturbation at horizon crossing time, $\delta_\mathrm{hc}$, at the time when $R_\mathrm{a} = H^{-1}_\mathrm{b}$, by solving for $\delta$ \Eq{Omega-delta relation} and substituting $\Omega$ from \Eq{Omega - sinxa relation}. Finally, one obtains that
\beq\label{delta at horizon crossing}
\delta_\mathrm{hc} = \left(\frac{H}{H_\mathrm{b}}\right)^2 - \cos^2\chi_\mathrm{a}.
\eeq
At the end,  on the uniform Hubble gauge, where $H=H_\mathrm{b}$, \Eq{delta at horizon crossing} becomes
\beq\label{delta at horizon crossing in the UH gauge}
\delta^\mathrm{UH}_\mathrm{hc} = \sin^2\chi_\mathrm{a},
\eeq 
We should stress out here that the above expression for $\delta^\mathrm{UH}_\mathrm{hc}$ was extracted independently on the equation of state of the universe at PBH formation time.

\subsection{Refining the PBH formation threshold}\label{sec:delta_c_refined}

We perform now a refined computation regarding the PBH formation threshold in the case of a time-dependent EoS parameter. In particular,  following the same reasoning as in~\cite{Harada:2013epa}, we compute $\delta_\mathrm{c}$ by establishing a criterion in order not to avoid the gravitational collapse. In particular, we confront the gravitational force which pushes the fluid matter of the collapsing overdensity inwards and enhances in this way the collapse to a black hole with the pressure gradient force which pushes matter outwards, thus delaying the collapsing process.  Practically, the criterion adopted is the requirement that the time at which the pressure sound wave crosses the radius of the collapsing overdensity should be larger than the free fall time from the maximum expansion to complete collapse.  Thus, the pressure gradient force will not have time to disperse the collapsing fluid matter to the background medium and prevent at the end the gravitational collapse process. 

To do so, we put firstly the Friedmann equation for the overdensity, namely \Eq{Friedmann equation - overdense region}, in a Tolman-Bondi form, valid for the case of dust matter, which has an analytic parametric solution.  To do so,  we redefine appropriately the scale factor $a$ and the cosmic time $t$ as follows:
\begin{eqnarray}\label{a and t transforms}
\tilde{a}=a e^{3I(a)} \label{a transform} \\
\mathrm{d}\tilde{t}=\mathrm{d}t e^{3I(a)} \left[1+3w(a)\right]\label{t transform} ,
\end{eqnarray}
where $I\left(a\right)\equiv\int_{a_\mathrm{ini}}^{a} \frac{w(x)}{x}\mathrm{d}x$ and the index $\mathrm{ini}$ denotes the initial time. Then, solving the continuity equation $\dot{\rho} + 3H(1+w)\rho = 0$ for a time-dependent EoS parameter and using the coordinate transformation of \Eq{a transform} and \Eq{t transform}, the Friedmann equation for the collapsing overdensity (\ref{Friedmann equation - overdense region}) can be recast in a dust form as
\beq\label{Friedmann equation - overdense region - dust form}
\left(\frac{\mathrm{d}\tilde{a}}{\mathrm{d}\tilde{t}}\right)^2=\frac{A}{\tilde{a}} - 1,
\eeq
where $A=\frac{\rho_\mathrm{ini}a^3_\mathrm{ini}}{3\Mp^2}$ and we have used the fact that $\frac{\mathrm{d}\tilde{a}}{\mathrm{d}\tilde{t}} = \frac{\mathrm{d}a}{\mathrm{d}t}$. The above equation once integrated gives the following parametric solution:
\beq\label{dust parametric solution}
\tilde{a}=\tilde{a}_\mathrm{max}\frac{1-\cos\eta}{2}, \quad  \tilde{t}=\tilde{t}_\mathrm{max}\frac{\eta-\sin\eta}{\pi},
\eeq
with $\eta\in \left[0,2\pi\right]$. In the above solution,  $\eta$ is the conformal time defined in terms of the redefined scale factor and cosmic time,  i.e. $\mathrm{d}\tilde{t}\equiv \tilde{a}\mathrm{d}\eta$ whereas $\tilde{a}_\mathrm{max}$ and $\tilde{t}_\mathrm{max}$ are the redefined scale factor and cosmic time at the maximum expansion time  respectively which read as:
\beq\label{a_max - t_max}
\tilde{a}_\mathrm{max}=\frac{\Omega_\mathrm{ini}}{\Omega_\mathrm{ini}-1}\tilde{a}_\mathrm{ini}, \quad \tilde{t}_\mathrm{max}=\frac{\pi}{2}\tilde{a}_\mathrm{max}.
\eeq

Concerning now the sound wave propagation in a close Friedmann geometry, the latter is dictated by the following equation:
\beq
a\frac{\mathrm{d}\chi}{\mathrm{d}t}=c_\mathrm{s}(t),
\eeq
where $c^2_\mathrm{s}$ is the sound speed of a perfect fluid with a time-dependent EoS parameter. In the case of an adiabatic perfect fluid,  $c^2_\mathrm{s}$ reads as [See also \App{app:sound speed calculation}]
\beq\label{c^2_s with w time dependent_main_text}
c^2_\mathrm{s}(\eta) = w(\eta) -\frac{1}{3\left[1+w(\eta)\right]\mathcal{H}(\eta)}\frac{\mathrm{d}w}{\mathrm{d}\eta},
\eeq
where one sees the presence of the time derivative of $w$ in the expression for the sound speed,  which makes $c^2_\mathrm{s}(\eta)$ different from its value when $w$ is constant, i.e. $c^2_\mathrm{s}(\eta)=w$.

Finally, making use of the conformal time definition $\eta$ the above equation can be recast as
\beq\label{sound wave equation}
\frac{\mathrm{d}\chi}{\mathrm{d}\eta}=\frac{c_\mathrm{s}(\eta)}{1+3w(\eta)}.
\eeq
One then can establish the PBH formation criterion as described before by demanding that the time at which the sound wave crosses the radius of the overdensity, i.e. $\eta(\chi_\mathrm{a})$ is larger than the time at which the overdensity reaches the maximum expansion, i.e. $\eta_\mathrm{max}=\pi$. In this way, the pressure gradient force will not have time to prevent the gravitational collapse whose onset time is considered here as the time of maximum expansion. To do so, in contrast with the treatment of ~\cite{Harada:2013epa}  one should solve numerically \Eq{sound wave equation} and demand that 
\beq\label{PBH formation criterion}
\eta_\mathrm{num}(\chi_\mathrm{a}) =\pi, 
\eeq
where $\eta_\mathrm{num}(\chi)$ is the numerical solution of  \Eq{sound wave equation} and $\chi_\mathrm{a}$ is the comoving scale at which the sound wave crosses the overdensity at the time of the maximum expansion. 
Consequently, from \Eq{delta at horizon crossing in the UH gauge}  the PBH formation threshold in the uniform Hubble gauge reads as 
\beq\label{delta_c w time dependent}
\delta^{\mathrm{UH}}_\mathrm{c}=\sin^2\chi_\mathrm{a},
\eeq 
with $\chi_\mathrm{a}$ being the solution of $\eta_\mathrm{num}(\chi_\mathrm{a}) =\pi$. 

At this point, one should stress out that the black hole apparent horizon should form after the onset of the gravitational collapse, i.e. the time of the maximum expansion. Thus, one should demand as well that $\eta_\mathrm{h}>\eta_\mathrm{max}=\pi$ where $\eta_\mathrm{h}$ is the time of formation of the apparent horizon which is obtained when $\frac{2M}{R} = 1$, where $M$ is the Misner-Sharp mass in spherical symmetric spacetimes [See in ~\cite{Misner:1964je, Hayward:1994bu} for more details]. A rigorous analysis shows that in the case of a closed FLRW universe, the condition $\frac{2M}{R} = 1$ gives that~\cite{Harada:2013epa}
\beq
\eta_\mathrm{h} =2\chi_\mathrm{a} \quad \mathrm{or} \quad 2\pi - 2\chi_\mathrm{a}.
\eeq
Given the fact that the coordinates in \Eq{overdense region metric} cannot cover entirely the overdense region of perturbations for which $\pi/2<\chi_\mathrm{a}<\pi$ we work here with perturbations for which  $0<\chi_\mathrm{a}<\pi/2$ and therefore $\eta_\mathrm{h}=2\pi - 2\chi_\mathrm{a}$. Demanding then that $\eta_\mathrm{h}>\eta_\mathrm{max}=\pi$ one has that $\chi_\mathrm{a}<\pi/2$. Here, one should point out that in the case $w$ is constant then $c^2_\mathrm{s}=w$ and \Eq{sound wave equation} can be solved analytically. In this regime,  solving the equation $\eta(\chi_\mathrm{a}) = \pi$ with $0<\chi_\mathrm{a}<\pi/2$ leads to the formula for $\delta_\mathrm{c}$ obtained in \cite{Harada:2013epa}.

One then can use the aforementioned semi-analytic scheme and apply it in the case of time-dependent $w$ epochs such the preheating epoch during which PBHs can be abundantly produced~\cite{GarciaBellido:1996qt, Hidalgo:2011fj,
    Suyama:2014vga, Zagorac:2019ekv,Martin:2019nuw} or the QCD phase transition where one also expects enhanced PBH production due to the softening of the EoS ~\cite{Jedamzik:1996mr,Carr:2019hud,Byrnes:2018clq}.

However, it is important to stress out that the prescription described above for the computation of $\delta_\mathrm{c}$ can be  only viewed as an approximate one since it requires the homogeneity of the central overdense core that is not the case when one is met with strong pressure gradients. It is valid then for situations in which $w\ll 1$. As noticed also in \cite{Musco:2018rwt,Escriva:2019phb}, the ``three-zone'' model initially introduced by \cite{Harada:2013epa} gives $\delta_\mathrm{c}$ for a very sharply peaked homogeneous overdensity profile, where the shape parameter $\alpha\rightarrow 0$,  but it does not take into account the shape dependence of the energy density profile discussed in \Sec{sec:intro} and the role of pressure gradients. These two effects can potentially disfavor the gravitational collapse and increase the value of $\delta_\mathrm{c}$.  In particular, broader energy density perturbation profiles with $\alpha\gg 1$ have the tendency to bounce back to the background medium and not collapse to a PBH. For this reason, the critical PBH formation threshold needs to be high enough in order for such perturbations to collapse. Consequently, the PBH formation threshold computed within the ``three-zone'' model can be viewed as a lower bound for $\delta_\mathrm{c}$.

\subsection{The PBH formation threshold in the comoving gauge}
Let us now express the PBH formation threshold in the comoving gauge which is the one mostly used in numerical simulations ~\cite{Musco:2004ak,Polnarev:2006aa,Musco:2008hv,Musco:2012au}. In the comoving gauge, the energy density perturbation at horizon crossing,  $\delta^\mathrm{com}_\mathrm{hc}$ can be written as~\cite{Musco:2005bua} 
\beq\label{delta_c - comoving gauge}
\delta^\mathrm{com}_\mathrm{hc}=Q(\eta)\frac{1}{3r^2}\frac{\mathrm{d}}{\mathrm{d}r}\left[r^3K(r)\right]r^2_\mathrm{m},
\eeq
where $r_\mathrm{m}$ is the comoving scale of the collapsing overdensity region, defined as the position of the maximum of the compaction function, i.e. $\mathcal{C}^{\prime}(r_\mathrm{m})=0$, $K(r)$ is the curvature profile in the quasi-homogeneous solution regime~\cite{Musco:2018rwt} and $Q$ is a function of time which is given by
\beq\label{Q}
Q(\eta) = 1 - \frac{H(\eta)}{a(\eta)}\int_{a_\mathrm{ini}}^{a(\eta)}\frac{\mathrm{d}a^\prime}{H(a^\prime)}.
\eeq
In the case of a constant equation of state, $Q= \frac{3(1+w)}{5+3w}$. For the case of the ``three-zone'' model considered here, $K(r)=1$ and $r_\mathrm{m}=\sin \chi_\mathrm{a}$ and as a consequence
\beq
\frac{1}{3r^2}\frac{\mathrm{d}}{\mathrm{d}r}\left[r^3K(r)\right]r^2_\mathrm{m} = \sin^2 \chi_\mathrm{a} = \delta^\mathrm{UH}_\mathrm{hc}.
\eeq
Therefore, the energy density perturbation at horizon crossing time in the comoving and the uniform Hubble gauge are related as follows
\beq\label{delta_com vs delta_UH}
\delta^\mathrm{com}_\mathrm{hc} = Q(\eta) \delta^\mathrm{UH}_\mathrm{hc}.
\eeq
\subsection{Limitations of the HYK prescription and further refinement}\label{sec:limitations_refinement}
In the prescription described above for the computation of $\delta_\mathrm{c}$ in a time-varying $w$ background, $\delta^\mathrm{UH}_\mathrm{hc}$ in \Eq{delta_com vs delta_UH} is computed by extracting the location in comoving coordinates where the sound wave crosses the collapsing overdensity at the time of maximum expansion, denoted here with $\chi_\mathrm{a,max}$.

Given the fact that within the ``three-zone'' model the conformal time range within which the gravitational collapse is completed is $\Delta \eta=2\pi$ as it can bee seen from \Eq{dust parametric solution},  the location $\chi_\mathrm{a}$ in the HYK prescription is uniquely fixed once and for all by the constant value of $w$ independently on the time variation of $w$ given that it is constant in time.  

However in the current regime,  given that $w$ varies with time,  one should subdivide the conformal time range in subranges where $\Delta \eta=2\pi$ and extract the location $\chi_\mathrm{a,max}$ for every time subrange by solving numerically the equation $\eta(\chi_a) = \pi$. This aspect is very important to be taken into account since in the case of a dynamical $w$ profile  $\chi_\mathrm{a,max}$ is not uniquely fixed by the value of $w$ once and for all but rather depends on the full time evolution of the EoS parameter.  At the end,  by computing the location $\chi_\mathrm{a,max}$ for every conformal time subrange one can make an interpolation between the values $\chi_\mathrm{a,max}$ computed within the different time subranges and extract $\chi_\mathrm{a,max}$  as a function of time. 
Consequently, $\delta^\mathrm{UH}_\mathrm{hc}$ will read as
\beq\label{eq:delta_c_refined_deltaeta}
\delta^\mathrm{UH}_\mathrm{hc} = \sin^2\left[\chi_{a,\mathrm{max}}(\eta)\right].
\eeq

At the end, this refined treatment takes into account the details of the full dynamical profile of the EoS parameter and constitutes a first step towards a more accurate computation of $\delta_\mathrm{c}$ in a time-varying $w$ background.  It also shed light on the limitations of the HYK prescription which can clearly not be used in a time-dependent $w$ regimes.

\section{The PBH formation threshold during the QCD phase transition}\label{sec:QCD}
We choose to illustrate our formalism regarding the computation of $\delta_\mathrm{c}$ with a rather physical regime observed in the universe where the EoS parameter changes with time.  One should in principle choose regimes where $w\ll 1$ in which the effect of pressure gradients is small an our formalism is precise.  Such a regime is the  phase of (pre)reheating where in the context of canonical inflation the universe effectively behaves as dust ($w\simeq 0$). However,  given that the details of the physics during reheating remain uncertain we do not have yet a robust answer of what is the dynamical evolution of $w$ during this phase~\cite{Munoz:2014eqa}.

For this reasom, we choose to illustrate our formalism with the case of  the QCD phase transiton where the EoS parameter and the sound speed are robustly computed through lattice QCD simulations.  During this period, one expects to have an enhanced PBH production with mass $M\sim M_\odot$ due to the softening of the equation of state~\cite{Jedamzik:1996mr,Carr:2019hud,Byrnes:2018clq}.  However,  during the QCD phase transition phase $w$ varies within the range $0.23<w<0.33$ where one expects a strong effect of the pressure gradients.  For this reason, the results presented in this section regarding the computation of $\delta_\mathrm{c}$ in a time-varying $w$ background should be viewed as an example case where our formalism is applied rather than a defined result of the paper.

\begin{figure}[h]
\centering
\includegraphics[width=0.495\textwidth]{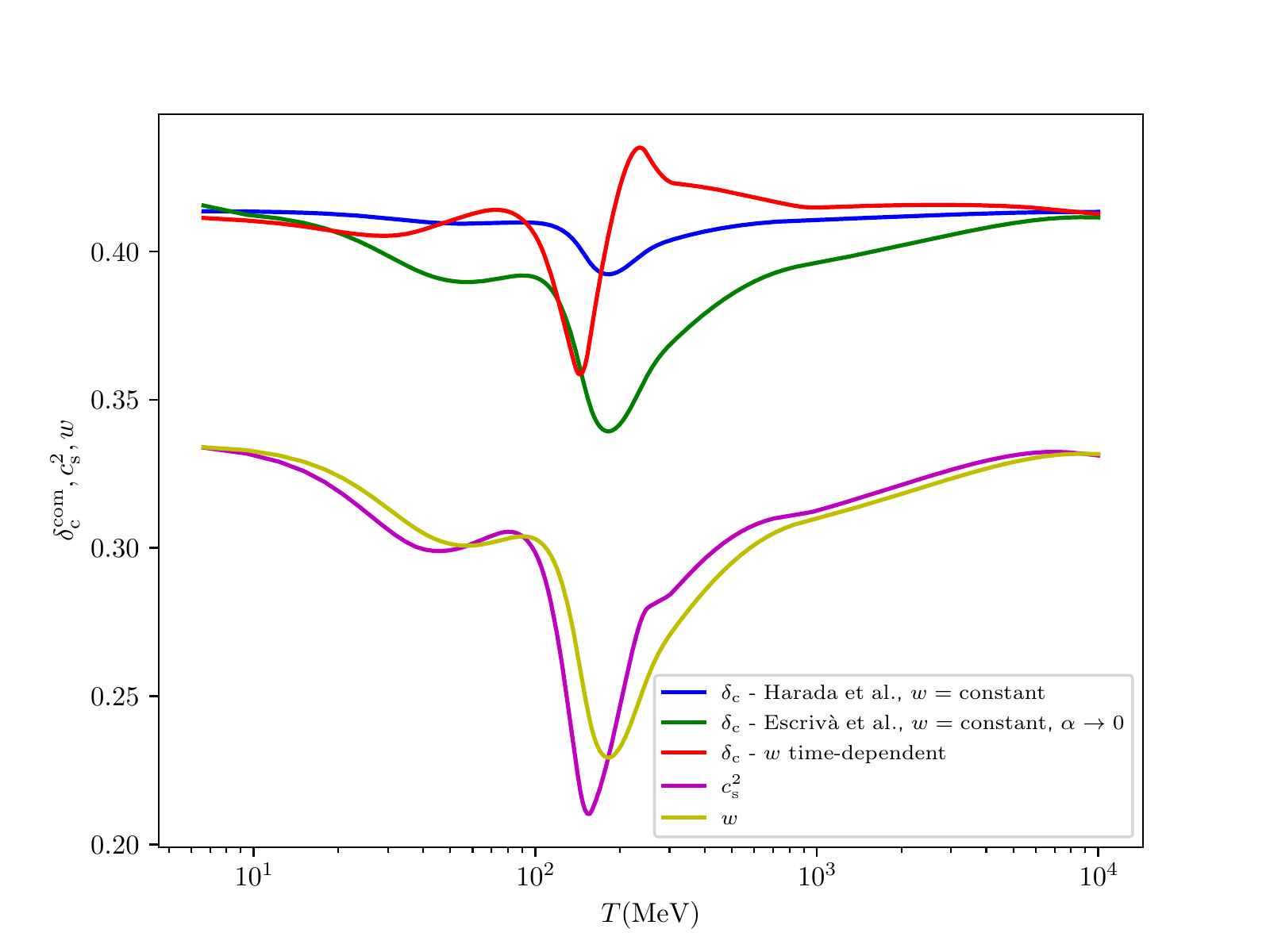}
\caption{In the bottom panel  we show the EoS parameter $w$ (yellow curve) and the sound speed square, $c^2_\mathrm{s}$ (magenta curve) as a function of the temperature during the QCD phase transition.  In the top panel,  we plot the PBH formation threshold, $\delta_\mathrm{c}$ in the comoving gauge during the QCD phase transition.  With the blue curve we depict $\delta_\mathrm{c}$ derived within the HYK prescription, valid for a constant EoS parameter $w$ while the green curve shows $\delta_\mathrm{c}$ computed within numerical simulations for constant $w$ by Escrivà et al. ~\cite{Escriva:2020tak} in the case of a peaked energy density perturbation profile with shape parameter $\alpha\rightarrow 0$. The red curve corresponds to $\delta_\mathrm{c}$ derived within our prescription where $w$ varies with time.}
\label{fig:QCD}
\end{figure}

Regarding now the period of the QCD phase transition, the EoS parameter $w$, defined as the ratio between the pressure and the energy density, i.e.  $w\equiv p/\rho$,  can be computed with the help of the number of energy and entropy relativistic degrees of freedom, $g_\rho$ and $g_s$ respectively, defined through the expressions for the energy and the entropy density, i.e $g_\rho(T) = 30\rho/(\pi^2T^4)$ and $g_s(T) = 45s/(2\pi^2T^3)$.  At the end, using the relationship between $\rho$, $s$ and $p$, i.e. $p=sT-\rho$, and the definition of $w$ one can easily find that the EoS parameter and the sound speed square read as:
\begin{align}
w(T) & = \frac{4g_s(T)}{3g_\rho(T)} - 1 \\
c^2_\mathrm{s}(T) & = \frac{4(4g_s(T)+Tg^\prime_s(T))}{3(4g_\rho(T)+Tg^\prime_\rho(T))} - 1.
\end{align}
Using now tabulated data for $T$, $g_\rho$ and $g_\rho/g_s$ during the QCD phase transition~\cite{Borsanyi:2016ksw} and making a cubic spline interpolation one can extract the EoS parameter $w$ and the sound speed square $c^2_s$ as a function of the temperature.  See in the bottom panel of \Fig{fig:QCD} their variation with temperature. 

Then,  applying the formalism presented before we compute $\delta_\mathrm{c}$ in the comoving gauge during the QCD phase transition (red curve in \Fig{fig:QCD}) by dividing the conformal time range of variation of $w$ in $N_\mathrm{p}=10^5$ subranges with $\Delta\eta = 2\pi$ following the procedure described in \Sec{sec:limitations_refinement}. By increasing  more the number of subranges, the picture does not change significantly. Then, we compare it with $\delta_\mathrm{c}$ obtained within the HYK treatment (blue curve in \Fig{fig:QCD}) valid for constant $w$ as well with the value of $\delta_\mathrm{c}$ obtained from numerical simulations (green curve in \Fig{fig:QCD}) with constant $w$ run by Escrivà $\&$ Germani $\&$  Sheth (EGS)~\cite{Escriva:2020tak} making at the end a linear interpolation between $w$ and $\delta_\mathrm{c}$ values in the regime where $w<1/3$. Regarding, the shape of the collapsing overdensity region we account for the fact that in our setup we have a peaked overdense region matched to the unperturbed background through an underdense layer.  Therefore, the shape parameter $\alpha$ introduced in \Sec{sec:intro} should be close to zero, i.e. $\alpha\rightarrow 0$.

As one may notice from \Fig{fig:QCD} , there are places in particular around the region of the minimum of $w$ or $c^2_\mathrm{s}$ where one is met with regimes where $c^2_\mathrm{s}\neq w$ and where $w$ varies within $30\%$ from its background radiation value $w_\mathrm{rad}=1/3$.  The case where $c^2_\mathrm{s}\neq w$ is actually the non static regime where in the context of a perfect adiabatic fluid, the EoS parameter $w$ varies with time as it can be seen from \Eq{c^2_s with w time dependent_main_text}. Clearly, in these regimes the range of validity of constant $w$ analytic or numerical prescriptions for the computation of $\delta_\mathrm{c}$ is limited.  

Interestingly,  one can also notice from \Fig{fig:QCD} that $\delta_\mathrm{c}$ follows the course of $c^2_\mathrm{s}$ with its minimum being situated at the same location with the minimum of $c^2_\mathrm{s}$.  One should also point out that in the regions where $c^2_\mathrm{s}\neq w$ our prescription differs significantly from the EGS and the HYK prescrtiptions where $c^2_\mathrm{s}=w$.  In particular, when $c^2_\mathrm{s}<w$, i.e. when the fluid becomes softer compared to the static case, $\delta_\mathrm{c}$ is decreased compared to the constant $w$ prescriptions enhancing in this way the PBH gravitational collapse whereas when $c^2_\mathrm{s}>w$, i.e. when the fluid becomes harder compared to the static regime, $\delta_\mathrm{c}$ is increased disfavoring the gravitational collapse.  Expectedly, when $c^2_\mathrm{s}\sim w$, $\delta_\mathrm{c}$ approaches its value computed within the EGS and the HYK prescrtiptions.

The above mentioned behavior can be explained also mathematically if one studies carefully the sound wave equation \eqref{sound wave equation} where one can see that the conformal time derivative $\frac{\mathrm{d}\chi}{\mathrm{d}\eta}$ is proportional to the ratio $\frac{c_\mathrm{s}(\eta)}{1+3w(\eta)}$.  So, in regions where $c^2_\mathrm{s}>w$, the conformal time derivative increases and as a consequence the location at which the sound wave crosses the overdensity region at the time of maximum expansion shifts to higher values of $\chi_{\mathrm{a,max}}$ always in the range $0<\chi_\mathrm{a}<\pi/2$ as mentioned in \Sec{sec:delta_c_refined}.  Consequently, given the fact that the $\sin^2$ function is a monotonically increasing function in the range $0<\chi_\mathrm{a}<\pi/2$ this leads to an increase of $\delta_\mathrm{c}$ as it can be clearly seen from \Eq{eq:delta_c_refined_deltaeta}. The inverse phenomenology is followed when $c^2_\mathrm{s}<w$ leading to a decrease of $\delta_\mathrm{c}$ compared to the static case.

One should also highlight here that as noticed in ~\cite{Escriva:2020tak} in the case of a constant $w$ the HYK analytic formula gives a lower bound for $\delta_\mathrm{c}$ in the regime where $w>1/3$,  i.e. when pressure gradients are not negligible. However, in the regime  $w<1/3$ there is a significant discrepancy between the HYK limit and the results from numerical simulations in the case of peaked energy density perturbation profiles, i.e. $\alpha\ll 1$, in contradiction with what it was believed up to recently, namely that the HYK formula gives a lower bound of $\delta_\mathrm{c}$.  This fact can be confirmed as well here since as one sees in \Fig{fig:QCD} in the region of the abrupt decrease of $w$ where $w<1/3$, $\delta_\mathrm{c}$ with the HYK prescription is higher of the order $10\%$ compared to the EGS results based on numerical simulations for constant $w$. 

However, within our prescription where the time variation of $w$ is taken into account we find indeed a lower bound for $\delta_\mathrm{c}$ at least in the regimes where the fluid becomes softer compared to the static case, i.e.  $c^2_\mathrm{s}<w$.  In the region, where the fluid becomes harder, i.e.  $c^2_\mathrm{s}>w$ the $\delta_\mathrm{c}$ value computed within our formalism is found to be higher compared to both the HYK and the EGS prescriptions.  

In particular,  there is a spike in $\delta_\mathrm{c}$ around $T\sim 200 \mathrm{MeV}$ which gives rise to a variation of the threshold of about $0.08$, a feature which is absent in previous results on the topic which assume however a constant EoS parameter.  This behavior may be explained by the fact that our prescription considers only the time variation of the background values of $w$ and $c^2_\mathrm{s}$ not accounting for the effect of the backreaction of the perturbations to these thermodynamic quantities which may alter our results. In principle,  one should consider the full perturbed $w$ and $c^2_\mathrm{s}$ and extract $\delta_\mathrm{c}$ from numerical simulations. According to ongoing numerical research work on the topic~\cite{RefToAdd}, this spike is not present when one includes the perturbations' backreaction effect and extract $\delta_\mathrm{c}$ from numerical simulations.  Thinking in physical terms,  this result can be explained from the fact that the pressure gradients constitute a form of gravitational energy so while initially they delay the gravitational collapse once the collapse is triggered they mostly favor it since as it was found in~\cite{Escriva:2020tak} the formation time of a black-hole apparent horizon is decreasing as $w$ increases.  In view of this physical understanding, the effect of the pressure gradient is expected to wash out the enhancement of $\delta_\mathrm{c}$ in the regions where $c^2_\mathrm{s}>w$ signaling that the respective spike in the $\delta_\mathrm{c}$ value in the region around $T\sim 110\mathrm{MeV}$ should not be interpreted as physical. 

\section{Conclusions}\label{sec:conclusions}
The PBH formation threshold $\delta_\mathrm{c}$ constitutes a very important quantity in the field of PBH physics since it determines the criterion for PBH formation and affects crucially the computation of the PBH abundace, a quantity which is constrained by many observational probes~\cite{Carr:2020gox}. The vast majority of the works in the literature has not accounted for the dynamical evolution of $w$ during the PBH formation process considering it as static, i.e. constant in time, an approximation which is good for the vast majority of the cosmic epochs. 

However,  in general $w$ is a time-dependent parameter and in particular in some periods of the cosmic history such as preheating or the QCD phase transition when one expects an enhanced PBH production, can experience very abrupt changes.  Consequently, motivated by the above mentioned phenomenology, in this work we studied the effect of a time-dependent EoS parameter on the computation of the PBH formation threshold $\delta_\mathrm{c}$ based on physical arguments. In particular,  generalising the mathematical formalism of~\cite{Harada:2013epa} and commenting its limitations we computed $\delta_\mathrm{c}$ by comparing the time the pressure sound wave crosses the radius of the collapsing overdensity with the time of the onset of the gravitational collapse. 

We should point out however that our refined semi-analytic treatment for the calculation of the PBH formation threshold  is an approximate one working well in regimes where pressure gradients are small.  Thus, the value of $\delta_\mathrm{c}$ derived here can be viewed as a lower bound of the true $\delta_\mathrm{c}$ at least in the regimes where the fluid becomes softer compared to the static case, namely when $c^2_\mathrm{s}<w$. It also does not take into account the dependence of $\delta_\mathrm{c}$ on the shape of the collapsing energy density perturbations as well as perurbation backreaction effects on the EoS parameter and the sound speed which may alter our conclusions.  

We applied our refined prescription for the computation of $\delta_\mathrm{c}$ in the case of the QCD phase transition, when the EoS parameter changes significantly with time and one expects an enhanced PBH production due to the abrupt softening of $w$.  Interestingly, we found that in the static regime,  where $c^2_\mathrm{s}\sim w$, which is mainly realised when $w$ and $c^2_\mathrm{s}$ varies slowly with time, our formalism gives results comparable with analytic and numerical approaches assuming a constant $w$.  

However, in the non static regime, i.e. regions where $c^2_\mathrm{s}\neq w$, our prescription differs significantly from prescriptions assuming constant $w$.  In particular,  by accounting only for the background dynamical evolution of $w$ and $c^2_\mathrm{s}$, when $c^2_\mathrm{s}<w$, i.e. when the fluid becomes softer compared to the static case, $\delta_\mathrm{c}$ is decreased compared to the constant $w$ prescriptions enhancing in this way the PBH gravitational collapse whereas when $c^2_\mathrm{s}>w$, i.e. when the fluid becomes harder, $\delta_\mathrm{c}$ is increased disfavoring the gravitational collapse.  

These interesting results show the big effect of a dynamical $w$ profile to the determination of the PBH formation threshold and points out to the need of the development of full numerical techniques to study the effect of a time-varying EoS parameter on the value of $\delta_\mathrm{c}$.

\section*{Acknowledgments}
I would like to thank Ilia Musco and Vincent Vennin for stimulating discussions and comments as well as Albert Escriv\`a for providing me with tabulated numerical data for $w$ and $\delta_\mathrm{c}$ in the regime of $w<1/3$ and $\alpha \rightarrow 0$.  I thank as well S. Clesse for bringing to my attention an erratum regarding the rescaling of the temperature x-axis of \Fig{fig:QCD}. I also acknowledge financial support from the Foundation for Education and European Culture in Greece.
\begin{appendix}
\section{The sound speed in a time-dependent $\textbf{$w$}$ background}\label{app:sound speed calculation}
Here, we extract the sound speed of a general adiabatic fluid, $c^2_\mathrm{s}$, with a time-dependent equation-of-state parameter, $w$. In a general system, the pressure density $p$ is a function of the energy density $\rho$ as well as of the entropy density $S$, i.e. $p=p(\rho,S)$. Consequently, one can write the following equation:
\beq\label{sound speed definition general}
\delta p =c^2_\mathrm{s}  \delta\rho + \left( \frac{\partial p}{\partial S}\right)_{\mathrm{\rho}} \delta S,
\eeq
where the sound speed $c^2_\mathrm{s}$ is defined as $c^2_\mathrm{s} \equiv \left(\frac{\partial p}{\partial\rho}\right)_{\mathrm{S}}$. If one considers then an adiabatic system then they should require that $\left(\frac{\partial p}{\partial S}\right)_{\mathrm{\rho}} = 0$, i.e. there is no entropy production. Consequently, for such a system $c^2_\mathrm{s}$ becomes
\beq\label{c^2_s definition}
 c^2_\mathrm{s} =\frac{\delta p}{\delta \rho}, 
 \eeq
Given then the fact that the background pressure and energy densities of an adiabatic fluid system, $p$ and $\rho$ depend only on time, one can rewrite \Eq{c^2_s definition} by introducing the derivation with respect to the conformal time using the chain rule, as
\beq
c^2_\mathrm{s} = \frac{p^\prime}{\rho^\prime},
\eeq
where the prime denotes differentiation with respect to the conformal time, $\eta$.  Finally, by using the continuity equation  $\rho^\prime + 3\mathcal{H}(1+w)\rho = 0$, written with the conformal time as the time variable as well as time differentiating $w$ defined as $w\equiv p/\rho$, one can straightforwardly obtain that
\beq\label{c^2_s with w time dependent}
c^2_\mathrm{s}(\eta) = w(\eta) -\frac{1}{3\left[1+w(\eta)\right]\mathcal{H}(\eta)}\frac{\mathrm{d}w}{\mathrm{d}\eta} ,
\eeq
where $\mathcal{H}\equiv \frac{a^\prime}{a}$ is the conformal Hubble parameter. 
\end{appendix}

\bibliography{PBH_w_t}
\end{document}